\documentclass[11pt,twoside]{article}

\usepackage{asp2006}
\usepackage{fancyhdr,graphicx,caption,rotating,multirow,lscape}
\DeclareGraphicsExtensions{.eps,.eps.gz,.ps,.jpg,.tiff}

\begin{document}

%%% Fill in title
\title{Detecting false alarms in transit data from space: Rejection methods tested in Corot Blind Test 2}

\author{J. M. Almenara,\altaffilmark{1} H. J. Deeg,\altaffilmark{1} C. R\'{e}gulo \altaffilmark{1} and R. Alonso\altaffilmark{2,1}}

\affil{$^{1}$Instituto de Astrof\'{i}sica de Canarias, 38200 La Laguna, Tenerife, Spain}
\affil{$^{2}$Laboratoire d'Astrophysique de Marseille, Traverse du Siphon, 13376 Marseille 12, France}

%\altaffiltext{1}{Instituto de Astrof\'{i}sica de Canarias, 38200 La Laguna, Tenerife, Spain}
%\altaffiltext{2}{Dpto. de Astrof\'\i sica, Universidad de La Laguna, La Laguna, 38206, Tenerife, Spain}
%\altaffiltext{3}{Laboratoire d'Astrophysique de Marseille, Traverse du Siphon, 13376 Marseille 12, France}

%\author{}
%\affil{}

\begin{abstract} %%% Abstract to run on from here.
Transit searches provide a large number of planet candidates. Before attempting follow-up observations, the best effort should be spent in classifying the light-curves, rejecting false alarms and selecting the most likely ones for real planets. A number of analysis tools has been developed with these objectives. Here, we apply such tools to 237 simulated multi-color light-curves from CoRoT Blind Test 2, which contain simulated planet transits and several configurations of impostors. Their comparison gives indications of the various tools'  classification and false-alarm rejection capabilities.  In order to arrive at the candidate identifications, we used an automated scheme of weighted punctuations assigned to the individual tests, which avoids that results from a single test dominate a candidate's classification.
\end{abstract}

%%% MAIN BODY OF TEXT GOES HERE. 
%%% Invited Talks : 10 pages
%%% Talks         : 6  pages
%%% Posters       : 3  pages
%%%IF YOU FEEL THAT THE INFORMATION PROVIDED 
%%% IN THIS FILE IS NOT SUFFICIENT, CONSULT THE FILE aspauthor2006.pdf. IT
%%% CONTAINS "INSTRUCTIONS FOR AUTHORS USING LATEX2E MARKUP", 
%%% SECTIONS 2.3-2.6 FOR HELP WITH EQUATIONS, FIGURES, AND TABLES.

\section {The Corot Blind Test 2}
The objective of the CoRoT Blind Test 2  (BT2) has been an evaluation of the capacity of transit characterization algorithms to identify correctly the events that cause transit-like signals. For this test, a set of 237 lightcurves was generated by two teams (curves 0 to 99 by M. Ollivier \emph{et al.} at IAS/Paris,  and curves 100-236 by C. Moutou \emph{et al.} at LAM/Marseille). The curves reproduce CoRoT's three colour Band-passes, plus a white band that is the sum of the three colours, and the expected noise sources \citep{corotbook2006} with the best possible fidelity. Each curve contained some transit-like features originated from simulated planets (in 38 cases) or binary systems (in 202 cases; some curves containing two objects). Similar to the 'Blind Test 1' \citep{BT1}, several teams were formed whose task was to identify the objects hidden in these lightcurves, and whose nature was unknown to these teams. Here we describe the strategies that were employed by the Team at IAC.

\begin{figure}[!ht]
\centering
\includegraphics[height=6.65cm,width=13.4cm]{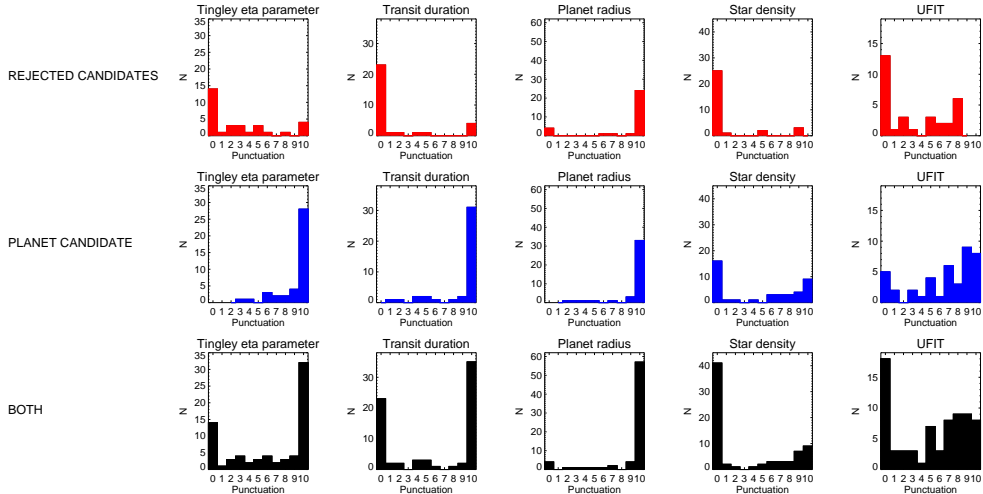}
\caption{Punctuations obtained for the candidates in the rejection tests of Sect. 2. These tests led to a split of the total sample (lower row) into 31 rejections (upper row) and 41 remaining planet candidates (middle row).\label{hist}}
\end{figure}
\section{Processing of the white lightcurves}   %%% Top level section head (remove "%" symbol)
First, a filter to remove spurious points and a high-pass filter was applied to clean the data. A transit search over the white light data was then performed, using the TRUFAS algorithm \citep{regulo2006}, employing a correlation with a transit-like wavelet function and a search for regularly spaced peaks. In 69 curves we could not identify events. In the remaining ones we then identified cases with different primary/secondary eclipses as binaries, using an automatic algorithm, finding 96 binaries. A series of rejection tests was then applied over the remaining 72 candidates. For these tests we assigned a 'planet-likeliness' on a scale of 0-10 points, were 0 indicates a 'false alarm' (Figure \ref{hist}). These tests are described as follows:
\begin{itemize}
\item {\bf{Tingley $\eta$ parameter}}. The $\eta_{\star}$ parameter by \citet{tingley2005} was developed to serve as a quick indicator for the planet-likelihood of a candidate, based on very simple assumptions. It gives an estimate of the ratio between an observed and a calculated transit duration. In difference to the published version - using a constant stellar radius-, we employed an estimated radius based on the (known) $T_{eff}$ and assuming Main-sequence stars. 
\item {\bf{Maximum transit duration test}}. From the radius of the star, the maximum possible duration of a transit can be estimated \citep[Eq. 3 in][]{seager2003}, assuming a planet of Jupiter-size. We compared this upper limit with the observed transit duration obtained from a trapezium fit. \emph{This test led to most of the rejections.}
\item {\bf{Radius of transiting planet candidate}}. We estimated the minimum size of the transiting planet from the drop in brightness, and again using the estimated stellar radius. Rejection occurred for radii larger than 2 $R_{Jup}$. Four candidates were rejected solely from this test.
\item {\bf{Stellar Density}}. Based on a trapezium fit to the data, we derived the stellar density \citep[Eq. 9 in][]{seager2003}, and compared with tabulated stellar densities for a given $T_{eff}$. This works well for cases with high signal to noise ratios, but else is unreliable, being very dependent on a precise estimate of the transit-duration. 
\item {\bf{Fit of planet model to light-curve}}. With initial planet parameters (e.g. inclination and planet-size) from the trapezoidal fit, stellar radius and mass based on $T_{eff}$, and limb-darkening values based on CoRoT passbands (provided by C. Barban), we attempted to adjust a model of a planetary transit with the transit fitting program UFIT \citep{deeg2006}. We evaluate if a good adjustment to the data can be reached while maintaining  planet-like parameters. This test was intended to detect curves with V-shaped (or otherwise 'un-planetlike') features, though we note that the current rejection criterion might still have room for improvement (some V-shaped cases were later rejected visually).
\end{itemize}
\begin{figure}[!h]
\centering
\includegraphics[angle=0,width=9cm]{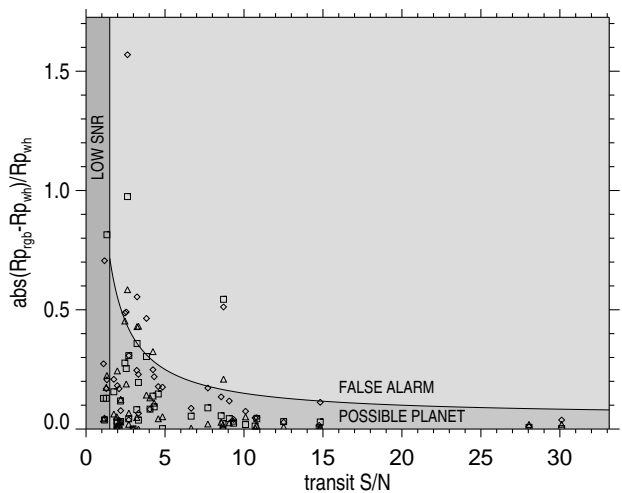}
\caption{Rejection from analysis of colored curves. The Y axis gives the normalized difference between red, green or blue, and white radii. A candidate is rejected if any of the 3 radius-differences falls within the zone marked as 'False Alarm'. For cases with a white light S/N $<$ 1.5, the S/N is too low to obtain a meaningful rejection criterion.\label{color}}
\end{figure}
\section{Analysis of colors of remaining candidates}
A weighted combination of the previous tests was used to define a sample of 41 planet candidates on which the color lightcurves were analyzed, using again the transit fitter UFIT. We first fitted the transit against white data and used this model as input to fits against each of the color curves, leaving only the planet radius as free parameter. Relevant differences between the white-light and the color-radii indicate grazing or diluted binary stars. The rejection of candidates is then based on a criterion that is a combination of the radius-comparison and the S/N of the candidate (Figure \ref{color}). This led to 29 planet candidates having passed the color rejection, 4 cases with a S/N too low (left zone in Fig. 2) for the test to be indicative, and 8 rejections. 
\section{Results from the tests}
The Tingley $\eta$ parameter and the maximum transit duration tests have been most powerful in finding false alarms, with a total of 23 rejections (out of the 72 candidates that were fed into this chain). The planet radius test has rejected only five candidates, but is important, as four of them were only recognized by this test. Further two tests, the planet-model fit and the stellar density comparison, were given low weight, providing reliable results only for high S/N. We also tried several other tests, which weren't used in the final candidate selection. These were the original versions of Tingley \& Sackett's (2005) $\eta$ parameters ($\eta_{p}$ and $\eta_{\star}$ with constant stellar radii), and a comparison of the transit durations obtained from planet-model-fitting against the maximum permissible transit durations. We also attempted to recognize the horn-like features predicted by \citet{tingley2004} in transit-curves of color-differences. However, it became apparent that these features will be recognizable only in data with much higher S/N than those of Blind Test 2, and consequently, those of CoRoT.

\begin{table}[!ht]
\caption{Summary of the detections.\label{summary_tab}}
\smallskip
\begin{center}
{\small  
\begin{tabular}{cccc}
\tableline
\noalign{\smallskip}
%\multicolumn{2}{c}{} & \multicolumn{4}{c}{Our Analysis}  \\
%\noalign{\smallskip}
%\cline{3-6}
%\noalign{\smallskip}
Simulated Input &Id. as planet &  Id. as binary  &  not detected \\
\noalign{\smallskip}
\tableline
\noalign{\smallskip}
38 Planets & \bf{18} & 1& 19  \\
202 Binaries & 10 & \bf{139}& 53 \\
%& Total & 28+5 & 72 & 135 & 240 \\
\noalign{\smallskip}
\tableline
\end{tabular}
}
\end{center}
\end{table}

The analysis described up to here has been made in a semi-automatic manner. In a real situation, the remaining 33 planet candidates would undergo an individual in-depth analysis. We only performed a visual revision, leading to 5 further rejections due to V-shaped lightcurves, resulting in a \emph{final count of 28 planet candidates} (see Table 1; correct identifications are in bold). Once the content of the simulations was revealed (Moutou \emph{et al.}, in prep), 10 candidates turned out to be false alarms. These were mainly cases were our rejection for differences among primary and secondary eclipses was set too conservative. However, we note that we followed in general a conservative rejection strategy, and only one planet got falsely identified as binary. This shows that a careful application of such rejection tests is a useful strategy to reduce a relatively large sample of transit candidates to one that is much more manageable for any further follow-up observations.

\end{document}